# Diagnostics of GNSS-based Virtual Balise in Railway Using Embedded Odometry and Track Geometry

Heekwon No, Jérémy Vezinet, Carl Milner, *Ecole Nationale de l'Aviation Civile*

**BIOGRAPHY (IES)**

Heekwon No is a postdoctoral researcher in the Ecole Nationale de l'Aviation Civile (ENAC), France. He received B.S. and Ph.D. degrees from the School of Mechanical and Aerospace Engineering at Seoul National University, South Korea. His research interests are the multi-sensor fusion based navigation, guidance and control system of the unmanned aerial or ground vehicle.

Jérémy Vezinet graduated as electronics engineer in 2010 and obtained his PhD in 2011 on multi-sensor hybridization from the ENAC (French National School for Civil Aviation) in Toulouse, France. Since 2011, he is a researcher and worked on multi-sensor hybridization research activities. He is now currently working on the use of GNSS for railway applications.

Carl Milner is an Assistant Professor within the Telecom Lab at the Ecole Nationale de l'Aviation Civile. He has a masters degree in Mathematics from the University of Warwick, a PhD in Geomatics from Imperial College London and has completed the graduate trainee programme at the European Space Agency. His research interests include GNSS augmentation systems, integrity monitoring, air navigation and applied mathematics.

**ABSTRACT**

The use of GNSS in the railway sector has been postulated on the notion of a Virtual Balise (VB). The VB-based positioning system works by setting a VB point on the railway track and determining the passage of the VB point using the position solution from the GNSS receiver. Although augmentation systems such as SBAS or GBAS are able to satisfy the integrity requirements of the aviation standards down to the $10^{-7}$ level, it is difficult to satisfy the high integrity requirements of the railway sector because firstly the railway users located on the ground are affected by the ground environment such as terrain, buildings and tunnels and secondly because the stringency of the railway sector requirements extends below the $10^{-9}$ level.

This paper proposes a method to detect faults in the GNSS solution due to satellite failure or local effects. Firstly, requirements for the monitoring performance are carefully derived accounting for the specificities of GNSS, namely that the possibility of consecutive VB faults cannot be discounted. The second contribution of the paper is the proposed detection using both odometry and track geometry of the onboard system. This enables to monitor all three-dimensional solution error so that higher sensitivity for the fault detection can be achieved.

Simulations have been performed with both single and dual (GPS, GALILEO) solutions. It has been found that the combinations of metrics are able to achieve very small missed detection probabilities for mean failure rates from 5.0m/s down to 0.03m/s for most dual constellation geometries. The detection performance of the odometer implementation varied according to the heading of the train. On the other hand, when odometry and track geometry are used together, all the three-directional monitors can obtain stable results regardless of the heading.

**INTRODUCTION**

GNSS has been widely used in the aviation and automotive sectors since it was made available to civilian users. Whilst in the aeronautical domain stringent safety requirements are in place, the rail sector presents an entirely different challenge. Firstly, requirements are set at a Safety Integrity Level (SIL) 4 [1] relating to $10^{-9}$ per hour hazard rates and secondly the local environment is subject to greater interference and potentially severe multipath. Consequently, the introduction of GNSS to the railway sector for safety critical applications has not been forthcoming. However, as interest in the automation and efficiency of the railway system has recently increased, research on the introduction of GNSS has been actively carried out.

The use of GNSS in the railway sector has been postulated on the notion of a Virtual Balise (VB). Existing railway positioning systems are based on various trackside infrastructures including Physical Balises (PBs) installed in groups between the tracks and odometry installed on-board. However, due to the high cost of installing and maintaining the PB, a GNSS-based VB positioning performed within on-board system has been proposed to reduce the role of PBs. Importantly however, the use of VBs will enable backward compatibility with existing architectures supporting PBs to accelerate and simplify the adoption of VBs.

The VB-based positioning system works by setting a VB point on the railway track and determining the passage of the VB point using the position solution from the GNSS receiver, replacing the existing PB. The actual PB delivers a message depending upon the direction of travel at the time of passing. The VB therefore must support similar capabilities.

As mentioned above, high safety requirements are required in the railway sector. Augmentation systems such as SBAS or GBAS, which are widely used in the aviation sector, are able to satisfy the integrity requirements of the aviation standards down to the $10^{-7}$ per hour level. However, it is difficult to satisfy the high integrity requirements of the railway sector because firstly the signals employed by railway users located on the ground are affected by the ground environment such as terrain, buildings and tunnels and secondly because the stringency of the requirements extends below the $10^{-9}$ level. Furthermore, at such a safety integrity level, rail requirements must be met through the provision of multiple functions. It is for this reason that previous research has proposed the use of odometry to help verify the absence of GNSS faults [2]. There are several researches focused on improving position accuracy or integrity by introducing additional sensors or information such as a sky-view camera [3], inertial measurement unit [4] and track map [5]. Usage of new sensors which are not embedded in the original system requires additional cost and complexity.

This paper proposes a method to detect faults in the GNSS solution due to satellite failure or local effects in advance of improving positioning accuracy. Firstly, requirements are carefully derived accounting for the specificities of GNSS, namely that the possibility of consecutive VB faults cannot be discounted, which has not been fully considered in some previous studies [6]. Such an assumption leads to requirements for the detection, or diagnosis of a GNSS fault through odometry, in terms of the duration of hazardous and not just the probability of occurrence. This represents the first new contribution of the paper.

The second contribution of the paper is the proposed detection using both odometry and track geometry of the onboard system. Especially, track geometry is useful for monitoring the deviations both laterally (across track) and vertically and detecting for the presence of ranging faults that may also impact the along track solution. Exponentially weighted moving averages (EWMA) of the difference in relative position change between GNSS, odometry and track geometry are derived. Currently a bank of four such metrics are used with varying scale averaging parameter. The combination of metrics enables the monitoring solution to detect both quickly varying faults, such as jumps or sharp ramps as well as slowly varying faults which ultimately, are the more difficult to detect.

The paper is organized as follows. First, the high-level architecture is described, including the notion of a virtual balise. And the hazards related to the use of the virtual balise are defined, which presents a different picture to previous work. In next section, odometry functions for detecting these hazards are derived. The new picture of proposed method is demonstrated through simulation on various conditions in aspect of considered GNSS constellation, geometry of satellite and magnitude of faults. Finally, the conclusion of this paper will be made.

**VIRTUAL BALISE ARCHITECTURE**

The positioning function plays important role in the automation of train. The Localization Unit (LU) determines the position of the train and transmits it to the trackside infrastructure. Using this information, the control system decides whether a movement authority can be given or not to the train. Existing positioning function of European Train Control System (ETCS) computes the position of a train based on the use of trackside infrastructure including *Eurobalises* and odometry. The Eurobalise is a PB which is installed on railway track and transmits a signal when a train passes over in one intended direction. The transmitted signal contains identification and position information of the PB. PBs are installed consecutively along railway track with specific interval. Average distance of 2 km between PBs is usually adopted in Europe for high speed train up to 300 km/h. The odometry function accumulates traveled distance from recently passed PB and provides position while the train is located between PBs. The architecture of PB based positioning function is represented in green blocks in Figure 1. Rail network operators, train manufacturers and rail industry stakeholders are trying to find a way to utilize GNSS to reduce the cost of maintaining current positioning system without alternating its architecture. A VB is proposed to provide positioning function with GNSS. It is compatible with the architecture of the current PB based LU so that the current system can migrate to a partially VB based system without refitting of all subsystems. A VB is a certain point set on the railway instead of the installation of a PB. The LU determines the passage of VB point based on combined position solution form GNSS and odometry. Utilization of correction from augmentation systems is also considered. The architecture of VB is represented in red blocks in Figure 1. GNSS with SBAS or GBAS has successfully provided integrity which meets its standard for aviation. However, providing integrity of GNSS to the railway is challenging because its higher required level of integrity and the effect of ground environment. In the section, hazards of VB based positioning system are classified.

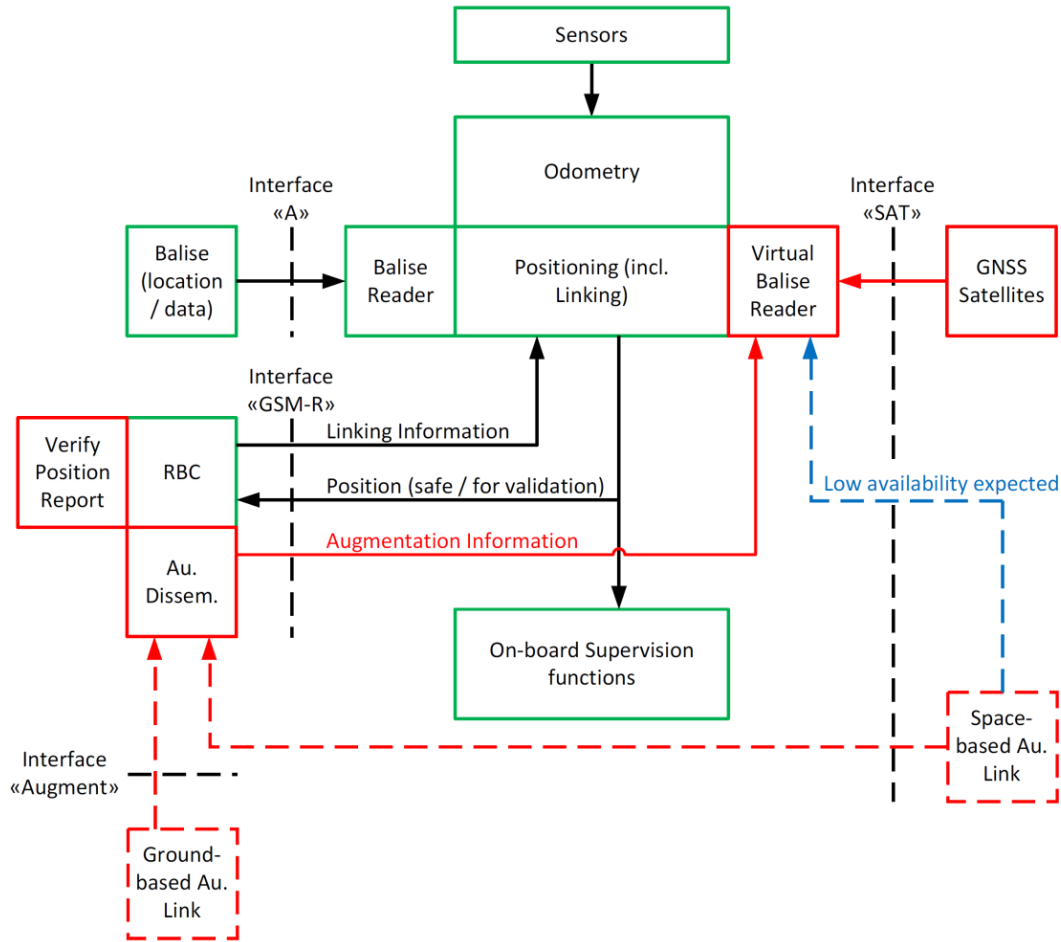

*Figure 1 Architecture of localization unit of train [7]*

**HAZARDS**

In the current PB based positioning architecture, three hazards were identified and are listed in [6]. These hazards are *corruption*, *deletion* and *insertion*. The corruption is when transmitted data from a PB contains an erroneous telegram, the deletion is when a PB over which train passes is not detected, and the insertion is when a train detects a PB that it should not (through cross-talk with a PB on a parallel track). The corruption does not need to be considered in VB based system since the transmission from VB will not actually occur and the telegram data will be stored in the LU on the train. Occurrence of a corruption or a deletion was monitored by comparing the order of IDs of detected balises with the expected order of those. Previous work [8] has specified the hazards of VB in same manner. However, in this paper, the nature of GNSS faults is used to derive the following classification.

*VB Jump* (VB-HAZARD 1) occurs when the position solution jumps from one epoch to the next thereby either crossing multiple virtual balises (VB-HAZARD 1A) or crossing in an excessive manner a single virtual balise (VB-HAZARD 1B).

*VB Shift* occurs when the position error drifts with time thus leading to virtual balise detection which is earlier or later than is correct (VB-HAZARD 2). In the extreme case, with a very high drift, this hazard is equivalent to (VB-HAZARD 1B).

VB Jump is, by definition, the least critical and least challenging. In the case of VB-HAZARD 1A, the checking order of detected balises enables diagnosis of a GNSS fault and the system passes to a fail-safe state. The second sub-case is subtler and requires an odometry based cross-check to diagnose its occurrence. Finally, for the VB shift case, the ability to diagnose such a fault will depend upon the degree of drift between the GNSS and odometry solutions, and its investigation is in the next section.

**DIAGNOSTIC MONITORS**

The basic notion of the diagnosis is that comparing the differential position from GNSS with that of odometry to detect fault of GNSS including a drift of position solution induced by a satellite failure or multipath. A raw monitor in the along-track direction at epoch k is introduced as follows:

$$q^{\Delta x}(k) = \Delta x^{gnss}(k) - \Delta x^{odo}(k) \tag{1}$$

$$\Delta x^{gnss}(k) = x^{gnss}(k) - x^{gnss}(k-1) \tag{2}$$

$$\Delta x^{odo}(k) = x^{odo}(k) - x^{odo}(k-1) \tag{3}$$

where $q^{\Delta x}$ is the raw monitor in the along-track direction, $x^{gnss}$ and $x^{odo}$ represent positions provided by GNSS and odometer in the along-track direction, respectively.

Assuming that the error components of GNSS and odometer position are zero-mean Gaussian, the statistic of the raw monitor is given by

$$q^{\Delta x}(k) \sim N(0, \sigma_{\Delta x}^2) \tag{4}$$

and variance of the raw monitor represented as

$$\sigma_{\Delta x}^2 = \sigma_{\Delta x, gnss}^2 + \sigma_{\Delta odo}^2 \tag{5}$$

where

$$\Delta x^{gnss}(k) \sim N(0, \sigma_{\Delta x, gnss}^2) \tag{6}$$

$$\Delta x^{odo}(k) \sim N(0, \sigma_{\Delta odo}^2) \tag{7}$$

For an SBAS-augmented single frequency code user, $\sigma_{\Delta x, gnss}$, the standard deviation of GNSS part is a meter-level and dominant. This raw monitor may be useful for detecting faults over a rate of several meter per second or jumps of several meters. However, it is not effective to detect faults of submeter per second level. Averaging techniques can be applied to the monitor to reduce its standard deviation and to improve detectability for slower faults. Exponential Weighted Moving Average (EWMA) is a sequential averaging technique and its level of averaging effect can be adjusted by setting its parameter of weighting decrease, $\alpha$. The EWMA applied monitor is given by

$$\bar{q}_\alpha^{\Delta x}(k) = \alpha \cdot q^{\Delta x}(k) + (1-\alpha) \cdot \bar{q}_\alpha^{\Delta x}(k-1) \tag{8}$$

The sequential form of EWMA gives great advantage of saving memory space when a long period of averaging window is required. The averaging effect of EWMA slows the response of monitor to the fault. Therefore, the following bank of monitors, which constitutes with the raw monitor and EWMA monitors with different $\alpha$ values, is used.

$$\begin{bmatrix} q^{\Delta x} \\ \bar{q}_{0.1}^{\Delta x} \\ \bar{q}_{0.01}^{\Delta x} \\ \bar{q}_{0.001}^{\Delta x} \end{bmatrix} \tag{9}$$

As odometry provides information of displacement in only the along-track direction, the sensitivity of fault detection can vary with the heading of train. To handle this issue, this paper proposes the usage of track geometry information. The position of the train is constrained to be located on the rail track. Therefore, the track geometry or map information can be used to derive a monitor in the cross-track and vertical direction. A ranging error in one specific direction affects the position error in the other directions as well as its direction, so that simultaneous monitoring of all three-dimension can improve the detectability. Raw monitors in the cross-track and vertical direction can be driven as follows by substituting the position information of odometry with that of the track geometry. The raw monitors in the cross-track and vertical direction are given as follows:

$$q^{\Delta y}(k) = \Delta y^{gnss}(k) - \Delta y^{map}(k) \tag{10}$$

$$q^{\Delta z}(k) = \Delta z^{gnss}(k) - \Delta z^{map}(k) \tag{11}$$

$$\Delta y^{gnss}(k) = y^{gnss}(k) - y^{gnss}(k-1) \tag{12}$$

$$\Delta z^{gnss}(k) = z^{gnss}(k) - z^{gnss}(k-1) \tag{13}$$

$$\Delta y^{map}(k) = y^{map}(k) - y^{map}(k-1) \tag{14}$$

$$\Delta z^{map}(k) = z^{map}(k) - z^{map}(k-1) \tag{15}$$

where $q^{\Delta y}$ and $q^{\Delta z}$ are the raw monitors in the cross-track and vertical direction, $y^{gnss}$ and $z^{gnss}$ represent positions provided by GNSS in the cross-track and vertical direction, $y^{map}$ and $z^{map}$ represent the cross-track and vertical components of projected GNSS position on railroad curve from the track geometry, respectively. The banks of monitors in the cross-track and vertical direction are also given by

$$\begin{bmatrix} q^{\Delta y} \\ \overline{q}^{\Delta y}_{0.1} \\ \overline{q}^{\Delta y}_{0.01} \\ \overline{q}^{\Delta y}_{0.001} \end{bmatrix}, \begin{bmatrix} q^{\Delta z} \\ \overline{q}^{\Delta z}_{0.1} \\ \overline{q}^{\Delta z}_{0.01} \\ \overline{q}^{\Delta z}_{0.001} \end{bmatrix} \tag{16}$$

With the banks of monitors in place, the thresholds for each monitor must be set as follows:

$$T = k_T \cdot \sigma \tag{17}$$

where $\sigma$ represents the standard deviation of a monitor, $T$ is the threshold corresponding to the monitor. The value for $k_T$ is taken corresponding to $10^{-7}$ of false alarm probability for each monitor. When at least one monitor value exceeds the corresponding threshold, a fault detection is made. The process of fault detection is represented in Figure 2. Note that the thresholds for each direction vary with the heading and satellite geometry. This variation of threshold is modeled with standard deviation of the along-track GNSS position by a linear regression method.

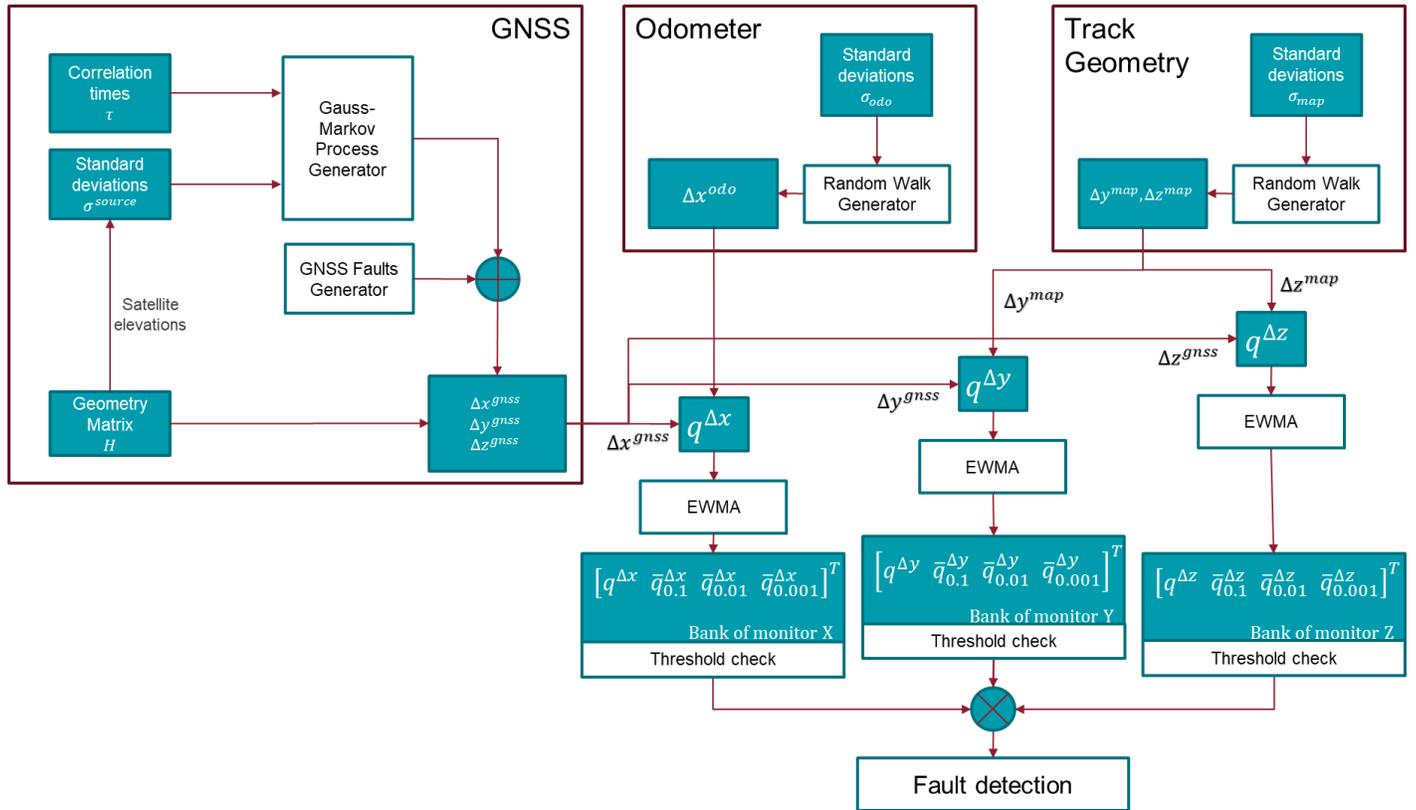

*Figure 2 Process of fault detection*

## ERROR MODELS

Modeling of GNSS errors follows previous works [9] and standard assumptions [10]. A Gauss Markov 1st order model is assumed for each error source including ionospheric error, tropospheric error, satellite orbit and clock error and the user error consisting of nominal multipath and noise [11]. The ionospheric error and the satellite orbit and clock errors are assumed to be remaining errors after application of SBAS correction. Details of GNSS error model are represented in Table 1.

*Table 1 GNSS Error model parameters*

| Error Source | $R(0)$ | $\tau$ | $\dfrac{R(\Delta = 10s)}{R(0)}$ | $\dfrac{R(\Delta = 1s)}{R(0)}$ |
|---|---|---|---|---|
| Units | $[m^2]$ | $[s]$ | | |
| Ionosphere | vary /w El [9] | 360s | 0.9726 | 0.9972 |
| Troposphere | vary /w El [9] | 1800s | 0.9945 | 0.9994 |
| Orbit/Clock | 0.3 | 3600s | 0.9987 | 0.9997 |
| User | 1.5 | 100s | 0.9048 | 0.9900 |

Two kinds of odometer, tachometer and radar are commonly equipped in train. The tachometer suffers from slipping and sliding when the train is accelerating and braking [2]. The radar shows consistent performance regardless of the movement of train. However, it can be erroneous when the facing ground surface is wet or covered with snow. Considering the described measurement uncertainty in [12] and assumed 1 Hz of measurement usage which is lower than raw output rate of sensor (>10 Hz), it is assumed that the combined odometry of tachometer and radar can give measurement with 0.05 m/s of noise.

Embedded track geometry data is assumed to give the cross-track and vertical coordinate of the train position with 1 m noise considering surveying error, track deformation error, and interpolation error along the surveyed points.

**SIMULATION**

Single constellation of GPS and dual-constellation of GPS and Galileo were tested in simulation to investigate the advantage of multi-constellation for fault detection. Range error was inserted into GPS satellite #8 for both settings of constellation. The inserted range error was ramp type, and the drift rate from 0.01 m/s to 5 m/s is tested. This range error accounts for a fault of satellite or a multipath error due to objects on the ground. Positioning failure is defined as the excess of 20 m of position error in the along-track direction.

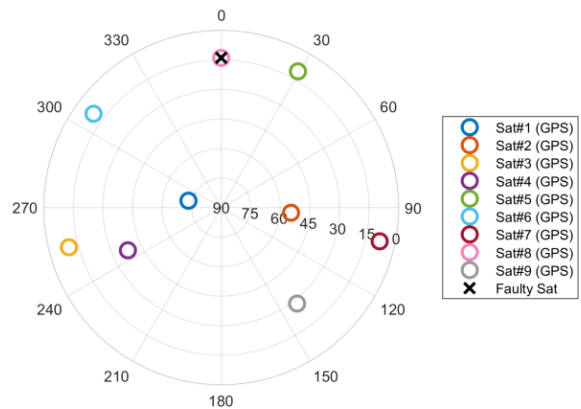

*Figure 3 Skyplot of GPS constellation*

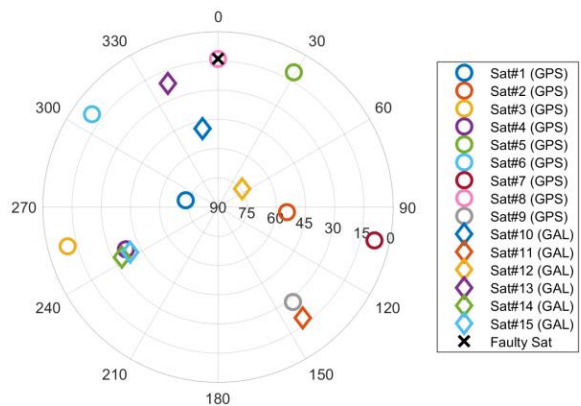

*Figure 4 Skyplot of dual-constellation of GPS and Galileo*

Monte Carlo simulations with $10^4$ repetitions were performed. Total time span of each simulation run was 15000 secs, and the range error was inserted from 5000 sec. The geometry of satellites and heading of the train were assumed to be fixed during simulation to consider a worst-case where the range error of slow drift rate can generate sufficiently large position error. Various heading of the train from 0 to 75 degrees with steps of 15 degrees are tested to confirm the effect of range error direction.

The results of the probability of missed detection ($P_{md}$) according to the time of detection since positioning failure ($T_{dsf}$) are shown in Figure 5. In the results, the heading angle is zero (the train heads toward the faulty satellite), and only the odometer is used for the diagnosis. The $T_{dsf}$ means elapsed time from the moment of position failure, when the position error in the along-track direction exceeds 20 m, and a negative value of it means that the fault is detected in advance of the occurrence of position failure. The results of $P_{md}$ for ramp failure of 0.03 m/s and 0.10 m/s show that the $P_{md}$ reaches zero when the $T_{dsf}$ has negative value. This means that the fault in range measurement is detected before the occurrence of position failure for all runs of simulation. The $P_{md}$ of dual-constellation decreases faster than that of GPS constellation so that the dual-constellation can detect the fault more quickly. However, in case of ramp failure of 0.01 m/s, $P_{md}$ cannot reach zero for both of constellation cases even after $T_{dsf}$ reaches positive value. This indicate that not all fault can be detected even after the occurrence of position failure.

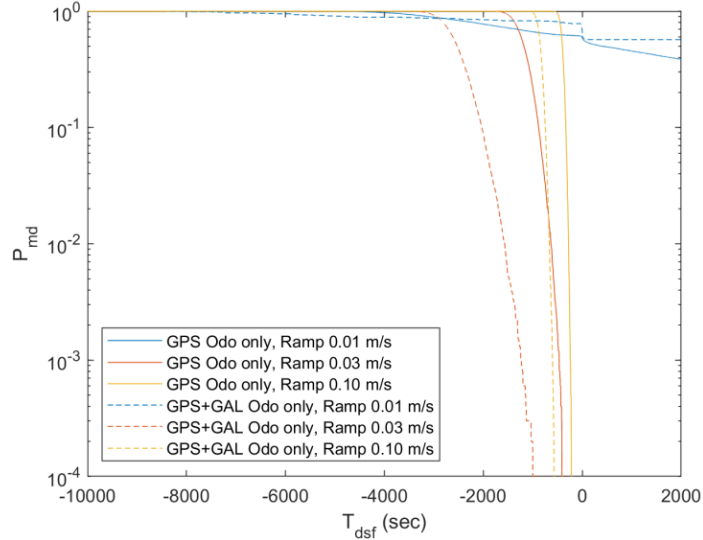

*Figure 5 Results of $P_{md}$ according to the time of detection since position failure*

The results of expected time of detection ($T_d$) according to variation of heading of the train are represented in Figure 6. The $T_d$ is the time taken for the detection since the occurrence of fault in range measurement. The results using only odometry show an increase of $T_d$ when the direction of faulty satellite is deviated from the heading of train since the odometry can provide only displacement information along advancing direction of the train. On the other hand, the track geometry can provide information for across track and vertical directions so that stable detection is possible when the odometry and the track geometry are used together.

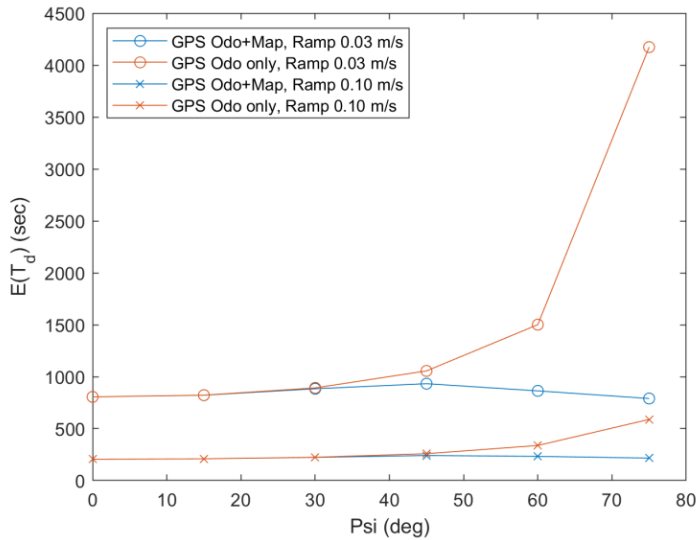

*Figure 6 Expected time of detection according to the heading of train*

The results of expected $T_{dsf}$ according to various heading of train and drift rate of failure are shown in Figure 7. The expected $T_{dsf}$ of dual-constellation leads that of GPS constellation case. The larger number of satellites in dual-constellation reduces the influence of fault on position solution, and the occurrence of position failure is postponed. As a result, greater margin of time between the detection of fault and position failure can be achieved.

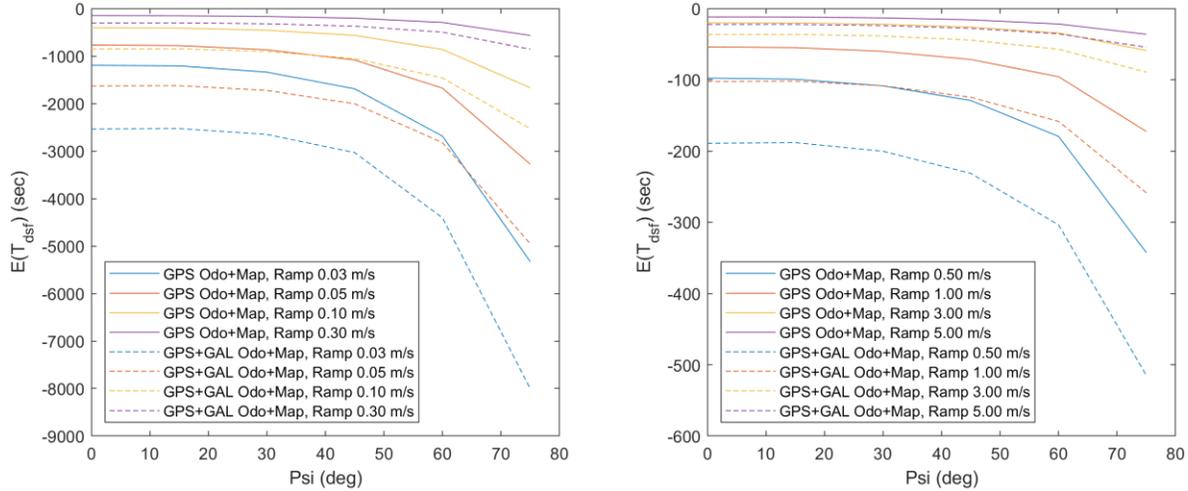

*Figure 7 Time of detection since position failure according to heading of train*

## CONCLUSION

A methodology of diagnosis of GNSS faults for train positioning system by utilizing embedded odometry and track geometry is proposed in this paper. The definitions of hazard suitable for GNSS based positioning system for train are suggested through considering the nature of GNSS faults. The diagnosis scheme comparing measurements of odometry and track geometry with position of GNSS is provided. The simulations were performed to investigate the effect of choice on GNSS constellation and the advantage of utilizing track geometry information. The simulation results are analyzed not only probabilistic manner but also time domain. The results show that the dual-constellation of GPS and Galileo is beneficial to secure sufficient margin of the detection time from the occurrence of position failure. The usage of track geometry improves consistency of the detection on the placement of faulty satellite or heading change of train. This paper suggests a sight to improve the integrity of railway by diagnostics before enhancing accuracy of positioning system using embedded sensor and track information.

## ACKNOWLEDGMENTS

This work has been funded by the ASTRail project. This project received funding from the Shift2Rail Joint Undertaking under the European Union's Horizon 2020 research and innovation programme under grant agreement No 777561.